\documentclass{article}
\usepackage{times}
\usepackage{amsfonts}
\usepackage{graphicx}
\begin{document}

\noindent
{\Large  A NOTE ON PADMANABHAN'S PATH--INTEGRAL\\ DUALITY}
\vskip1cm
\noindent
{\bf P. Fern\'andez de C\'ordoba}$^{1}$ and {\bf J.M. Isidro}$^{2}$\\
Instituto Universitario de Matem\'atica Pura y Aplicada,\\ Universitat Polit\`ecnica de Val\`encia, Valencia 46022, Spain.\\
$^{1}${\tt pfernandez@mat.upv.es}, $^{2}${\tt joissan@mat.upv.es}\\
\vskip.5cm
\noindent
\today
\vskip.5cm

\noindent
{\bf Abstract} We develop a field--theoretic implementation of Padmanabhan's principle of invariance under the duality $\mathcal S \rightarrow {L_0^2}/{\mathcal S}$, where ${\mathcal S}$ is the proper length (the action integral along a particle's trajectory) and $L_0$ is a quantum of length. Starting from the particle propagator and its Schwinger proper--time representation, we derive the wave operator associated with the duality transformation in terms of a scalar field, rather than a point particle. The resulting theory is governed by a nonlocal operator involving modified Bessel functions. We analyse the infrared and ultraviolet regimes, demonstrate the recovery of standard Klein--Gordon theory at low energies, and show the appearance of exponential ultraviolet suppression. Our analysis provides a natural extension, to quantum field theory, of Padmanabhan's duality for point particles.

\tableofcontents

\section{Introduction}

\subsection{Motivation}

One of the most robust expectations arising from virtually all approaches to quantum gravity is the existence of a fundamental length scale, or a quantum of length, $L_0$ (for a broad overview see, {\it e.g.}\/, ref. \cite{KIEFER}). The existence of a minimal length suggests that the classical notion of spacetime as a continuum may cease to be meaningful below a certain scale. Consequently, short--distance singularities and ultraviolet divergences are expected to be modified. A particularly attractive realisation of this idea was proposed by Padmanabhan \cite{PADDY1, PADDY2} through the principle of duality invariance. Rather than modifying the geometry directly, this proposal modifies the propagation amplitude associated with relativistic trajectories. This approach possesses several remarkable features. First, it effectively introduces a zero--point length $L_0$. Second, it preserves Euclidean invariance. Third, it regularises the coincidence limit of quantum propagators while requiring only a minimal modification of the relativistic particle action.

\subsection{The duality principle}

A free, massive scalar particle propagating on a Riemannian manifold ${\cal M}$ is described classically by an action integral 
\begin{equation}
m{\cal S}\left[p({\bf x}, {\bf y})\right]=m\int_{\bf y}^{\bf x}{\rm d}s,
\label{linint}
\end{equation}
the integral being computed along the path $p({\bf x}, {\bf y})$ connecting ${\bf y}\in{\cal M}$ to ${\bf x}\in{\cal M}$. Quantum--mechanically, the sum over all paths connecting ${\bf y}$ with ${\bf x}$, 
\begin{equation}
{\cal G}({\bf x}, {\bf y})=\sum_{{\rm paths}\,p}\exp\left\{-m{\cal S}\left[p({\bf x}, {\bf y})\right]\right\},
\label{treintayuno}
\end{equation}
defines the Euclidean propagator. In $\mathbb{R}^D$, it turns out to depend on the points ${\bf x}, {\bf y}$ through the geodesic distance between them. 

Padmanabhan's duality principle is the requirement of invariance under the transformation \cite{PADDY1, PADDY2}
\begin{equation}
{\cal S}\longrightarrow\frac{L_0^2}{\cal S}.
\label{sesenta}
\end{equation}
In order to achieve invariance one modifies the path integral (\ref{treintayuno}) to the manifestly duality--invariant path integral
\begin{equation}
{\cal G}_{\rm sd}({\bf x}, {\bf y})=\sum_{{\rm paths}\,p}\exp\left(-m\left\{{\cal S}\left[p({\bf x}, {\bf y})\right]+\frac{L^2_0}{{\cal S}\left[p({\bf x}, {\bf y})\right]}\right\}\right).
\label{treintaydos}
\end{equation}
The subscript ``sd" stands for {\it selfdual}\/. It has been established in refs. \cite{PADDY1, PADDY2} that the propagator (\ref{treintaydos}) includes lowest--order, quantum--gravitational effects due to the presence of  the zero--point  length $L_0$. 

Demanding invariance under the duality transformation (\ref{sesenta}) does not alter the underlying manifold structure. In what follows we will consider the case when the manifold ${\cal M}$ is Euclidean space $\mathbb{R}^D$, where $s^2({\bf x}, {\bf y})=\sum_{i=1}^D(x^i-y^i)^2$.

\subsection{Aim}

Padmanabhan's duality (\ref{sesenta}) is naturally formulated at the level of relativistic particle trajectories, whereas quantum field theory is formulated in terms of operator--valued fields. What is the Euclidean quantum field theory whose Green function reproduces the duality--modified propagator? 

The aim of the present work is to construct the corresponding field--theoretic description directly from the Euclidean propagator. We will show that the resulting theory is nonlocal and is governed by an operator involving modified Bessel functions. This operator emerges naturally from a Schwinger--like proper--time representation (see, {\it e.g.}\/, ref. \cite{PADDYQFT}) and requires no additional assumptions.

\section{The scalar propagator}

\subsection{Schwinger's representation}

The quantum theory of a real, massive scalar field in Euclidean $\mathbb{R}^D$ is described by the Klein--Gordon action\footnote{We stress the difference between the action integrals (\ref{linint}) and (\ref{ochenta}) by using the different fonts ${\cal S}$ and $S$. The corresponding Euclidean propagators are denoted ${\cal G}$ and $G$, respectively. The point--particle action ${\cal S}$ has mass dimension $-1$; the field--theory action $S$ is dimensionless.}
\begin{equation}
S\left[\Phi\right]=\frac{1}{2}\int {\rm d}^D{\bf x}\,\Phi\left(-\nabla^2+m^2\right)\Phi.
\label{ochenta}
\end{equation}
The corresponding Euclidean propagator is represented by the path integral
\begin{equation}
G(s)=Z^{-1}\int D\Phi\,\Phi({\bf x})\Phi({\bf 0})\exp\left(-S[\Phi]\right),
\label{ochetauno}
\end{equation}
where
\begin{equation}
Z=\int D\Phi\exp\left(-S\left[\Phi\right]\right)
\label{trescientosdiez}
\end{equation}
is the partition function for the scalar field $\Phi$.  One has
\begin{equation}
{\cal G}(s)=\frac{m^{D-2}}{(2\pi)^{D/2}}\frac{K_{D/2-1}\left(ms\right)}{\left(ms\right)^{D/2-1}}=G(s).
\label{doscientoscuarentanueve}
\end{equation}
In particular, the point--particle propagator ${\cal G}(s)$ and the Klein--Gordon propagator $G(s)$ are equal \cite{PADDYQFT}. Above, $K_n(z)$ is a modified Bessel function.\footnote{In our use of special functions we follow the conventions of ref. \cite{GRADSHTEYN}.} 

Consider the positive, elliptic operator on $\mathbb{R}^D$
\begin{equation}
A := -\nabla^2 + m^2.  
\label{uno}
\end{equation}
The corresponding Green function satisfies $G(A)=A^{-1}$, which can be written as 
\begin{equation}
G(A)=A^{-1} = \int_0^\infty {\rm d}s\, {\rm e}^{-sA},
\label{dos}
\end{equation}
where the parameter $s$ is Schwinger's proper time. 
  
We now incorporate Padmanabhan's duality principle. The prescription is \cite{PADDY1, PADDY2}: the replacement 
\begin{equation}
\exp(-sA) \longrightarrow \exp\left( -sA-\frac{L_0^2}{4s} \right), 
\label{once}
\end{equation}
once made inside the Schwinger representation (\ref{dos}) of the Euclidean propagator, effectively implements the desired duality. Consequently the selfdual propagator becomes
\begin{equation}
G_{\rm sd}(A) = \int_{0}^{\infty} {\rm d}s \, \exp \left( -sA - \frac{L_0^2}{4s} \right).
\label{cuatro}
\end{equation}

\subsection{Evaluation of the selfdual propagator}

We now proceed to evaluate the field--theory propagator (\ref{cuatro}).

Since the operator (\ref{uno}) is positive and selfadjoint, the spectral theorem implies the existence of a spectral decomposition 
\begin{equation} 
A = \int_{\sigma(A)} \lambda\, {\rm d}E_A(\lambda),
\label{cinco}
\end{equation} 
where $E_A(\lambda)$ denotes the spectral measure of $A$, and $\sigma(A)$ the spectrum.\footnote{In the absence of boundary conditions other than fast decay of $\Phi$ at infinity, $\sigma(A)=[m^2,\infty)$.} Therefore 
\begin{equation} 
{\rm e}^{-sA} = \int_{\sigma(A)} {\rm e}^{-s\lambda}\, {\rm d}E_A(\lambda). 
\label{seisbis}
\end{equation} 
Substituting (\ref{seisbis}) into (\ref{cuatro}) yields 
\begin{equation} 
G_{\rm sd}(A) = \int_{\sigma(A)} {\rm d}E_A(\lambda) \int_0^\infty {\rm d}s\, \exp\left(-s\lambda-\frac{L_0^2}{4s}\right).
\label{seis}
\end{equation} 
Now we have 
\begin{equation} 
I(\lambda)=\int_0^\infty {\rm d}s\, \exp\left(-s\lambda-\frac{L_0^2}{4s}\right)= L_0 \lambda^{-1/2} K_1\left(L_0\sqrt{\lambda}\right) 
\label{siete} 
\end{equation} 
which, plugged into the spectral decomposition (\ref{seis}), yields
\begin{equation} 
G_{\rm sd}(A) = \int_{\sigma(A)} I(\lambda)\, {\rm d}E_A(\lambda)=L_0 \int_{\sigma(A)} \lambda^{-1/2} K_1\left(L_0\sqrt{\lambda}\right) \,{\rm d}E_A(\lambda).
\label{ochobis}
\end{equation} 
One easily recognises this as the operator
\begin{equation} 
G_{\rm sd}(A) = L_0 A^{-1/2} K_1\left(L_0\sqrt A\right). 
\label{exact_propagator}
\end{equation} 
 
For later applications, it is useful to reexpress the above in momentum space, where $A$ becomes $p^2+m^2$. Then Eq. (\ref{exact_propagator}) becomes 
\begin{equation} 
G_{\rm sd}(p) = \frac{L_0} {\sqrt{p^2+m^2}} K_1 \left( L_0\sqrt{p^2+m^2} \right).
\label{Gmomentum} 
\end{equation} 
Eqs. (\ref{exact_propagator}) and (\ref{Gmomentum}) implement Padmanabhan's duality in terms of a scalar field. We see that the structure of the propagator differs substantially from the ordinary Klein--Gordon case.

\section{Properties of the selfdual propagator}

Next we analyse the properties of the momentum--space propagator (\ref{Gmomentum}). Our goal is to understand how the replacement (\ref{once}) modifies the standard Klein--Gordon theory.

\subsection{Infrared behaviour}

The infrared regime corresponds to $L_0\sqrt{p^2+m^2} \ll 1$. For small arguments, the function $K_1(z)$ admits the expansion
\begin{equation} 
K_1(z) = \frac{1}{z} + \frac{z}{2} \left( \ln\frac{z}{2} + \gamma - \frac12 \right) + \ldots, 
\label{K1IR} 
\end{equation} 
where $\gamma$ is Euler's constant. Use of (\ref{K1IR}) in Eq. (\ref{Gmomentum}) reproduces the Klein--Gordon propagator:
\begin{equation} 
\lim_{L_o\to 0}G_{\rm sd}(p) = \frac{1}{p^2+m^2}=G(p).
\label{KGpropagator} 
\end{equation} 
We conclude that the replacement (\ref{once}) leaves the infrared behaviour unchanged. The effects of the fundamental length scale \(L_0\) become visible only through subleading corrections. Eq. (\ref{KGpropagator}) constitutes an important consistency check.

\subsection{Ultraviolet behaviour}

The ultraviolet regime corresponds to $L_0\sqrt{p^2+m^2} \gg 1$. For large arguments, the function $K_1(z)$ admits the asymptotic expansion 
\begin{equation} 
K_1(z) = \sqrt{\frac{\pi}{2z}} \,e^{-z} \left( 1+ \frac{3}{8z} + \ldots \right). 
\label{veintiuno} 
\end{equation} 
Substituting the above into Eq. (\ref{Gmomentum}) one arrives at 
\begin{equation} G_{\rm sd}(p) = \sqrt{\frac{\pi L_0}{2}} \, \left(p^2+m^2\right)^{-3/4} \, \exp \left( -L_0\sqrt{p^2+m^2} \right) +\ldots  
\label{UVpropagator} 
\end{equation} 
We see that the selfdual propagator exhibits exponential UV damping. This behaviour differs radically from that of the ordinary Klein--Gordon propagator. Consequently the replacement (\ref{once}) naturally implements an effective UV cutoff without introducing a discrete spacetime structure. This remarkable property suggests that Padmanabhan's duality principle may provide a physically motivated mechanism for ameliorating UV divergences while preserving the continuity of the underlying spacetime manifold.

\section{The selfdual wave operator}

Padmanabhan's duality principle defines a consistent modification of ordinary scalar field theory. We now proceed to construct the wave operator whose Green function is given by equation (\ref{Gmomentum}).

\subsection{Definition and spectral decomposition}

Eq.  (\ref{dos}) implies that the Klein--Gordon operator $A$ and the corresponding Euclidean propagator $G(A)$ are mutually inverse. Such was the case prior to imposing  duality invariance. We now have the selfdual Euclidean propagator $G_{\rm sd}$ at our disposal. It thus makes sense to define a selfdual wave operator $A_{\rm sd}$ through the requirement 
\begin{equation} 
A_{\rm sd}\, G_{\rm sd} = {\bf 1}, 
\label{inversecondition} 
\end{equation} 
where ${\bf 1}$ denotes the identity. By Eq. (\ref{exact_propagator}) one immediately obtains 
\begin{equation} 
A_{\rm sd} = \frac{\sqrt A} {L_0K_1\left(L_0\sqrt A\right)}  
\label{exactK} 
\end{equation} 
as the duality--invariant wave operator. Eq. (\ref{exactK}) provides a field--theoretic realisation of Padmanabhan's duality principle. 

The operator (\ref{exactK}) is best understood through its spectral decomposition:
\begin{equation} 
A_{\rm sd} = \int_{\sigma(A)} \frac{\sqrt{\lambda}} {L_0K_1\left(L_0\sqrt{\lambda}\right)} \,{\rm d}E_A(\lambda). 
\label{spectralK} 
\end{equation} 
So the requirement of selfduality enforces the replacement of the spectral function $f(\lambda)=\lambda$ in (\ref{cinco}) with its selfdual counterpart
\begin{equation} 
f_{\rm sd}(\lambda) = \frac{\sqrt{\lambda}} {L_0K_1\left(L_0\sqrt{\lambda}\right)}. 
\label{quince}
\end{equation} 
Since $\sigma(A)>0$ and $K_1(z)>0$ whenever $z>0$, it follows that $f_{\rm sd}(\lambda)>0$. Consequently the selfdual wave operator $A_{\rm sd}$ is positive. It is also selfadjoint. 

Unlike the ordinary Klein--Gordon operator, equation (\ref{exactK}) is not a polynomial in the Laplacian $\nabla^2$. Instead it contains a modified Bessel function of the  operator $\sqrt{-\nabla^2+m^2}$.  The resulting theory is therefore nonlocal. The duality principle therefore leads naturally to a nonlocal field theory without requiring any additional assumptions beyond the existence of a fundamental length scale.
 
Eq. (\ref{quince}) allows one to investigate the IR and the UV behaviour of the selfdual field operator. This we do next.

\subsection{Infrared expansion}

Here the relevant regime is $L_0\sqrt{\lambda} \ll 1$. Then the expansion (\ref{K1IR}), applied to Eq. (\ref{quince}),  yields the corresponding spectral function
\begin{equation} 
f_{\rm sd}(\lambda) = \lambda - \frac{L_0^2}{2} \lambda^2 \left( \ln\frac{L_0\sqrt{\lambda}}2 + \gamma - \frac12 \right) + \ldots
\label{KspectralIR} 
\end{equation} 
By the spectral theorem, this translates into 
\begin{equation} 
A_{\rm sd} = A - \frac{L_0^2}{2} A^2 \left( \ln\frac{L_0\sqrt{A}}2 + \gamma - \frac12 \right) + \ldots
\label{KoperatorIR} 
\end{equation} 
The leading term reproduces the ordinary Klein--Gordon operator as expected. The first correction appears at order $L_0^2$ and contains the nonanalytic term $A^2\ln A $.

Unlike conventional effective field theories, where corrections are typically expressed as power series in $A=-\nabla^2+m^2$, the selfdual operator contains a logarithmic dependence on the spectrum. This feature reflects the intrinsically nonlocal nature of the theory. Moreover, the appearance of $A^2\ln A$ suggests that Padmanabhan's  duality principle cannot be interpreted simply as a local, higher--derivative modification of Klein--Gordon theory. Rather, it defines a new class of spectral operators whose ultraviolet behaviour is ultimately controlled by the modified Bessel function $K_1(z)$.

\subsection{Ultraviolet expansion}

Here the relevant regime is $L_0\sqrt{\lambda} \gg 1$. Applying the expansion (\ref{veintiuno}) to Eq. (\ref{quince}) yields the corresponding spectral function
\begin{equation} 
f_{\rm sd}(\lambda) = \sqrt{\frac{2}{\pi L_0}} \,\lambda^{3/4} \exp \left( L_0\sqrt{\lambda} \right)  +\ldots
\label{UVoperatorSpectral} 
\end{equation} 
We see that the ultraviolet growth of the wave operator is exponential. In momentum space, where $\lambda=p^2+m^2$, the above
becomes 
\begin{equation} 
A_{\rm sd}(p) = \sqrt{\frac{2}{\pi L_0}} \, (p^2+m^2)^{3/4} \exp \left( L_0\sqrt{p^2+m^2} \right) +\ldots
\label{UVoperatorMomentum} 
\end{equation} 
The selfdual operator $A_{\rm sd}$ therefore diverges exponentially in the UV. This growth is precisely responsible for the exponential suppression exhibited by 
$G_{\rm sd}$. By contrast, the ordinary Klein--Gordon operator $A(p) = p^2+m^2$ grows only polynomially at high momenta. 

From this viewpoint, the fundamental length scale $L_0$ acts as a spectral regulator. Rather than eliminating ultraviolet modes through a hard cutoff, the theory increases their effective kinetic energy exponentially. The resulting suppression is smooth, Euclidean invariant, and entirely determined by the duality principle.

\subsection{The selfdual action}

{}Finally we may formulate the scalar field theory based on the selfdual wave operator (\ref{exactK}). One simply replaces the action integral (\ref{ochenta}) with its selfdual counterpart
\begin{equation} 
S_{\rm sd}\left[\Phi\right] = \frac{1}{2} \int {\rm d}^D{\bf x}\, \Phi\, A_{\rm sd}\, \Phi, 
\label{treinta} 
\end{equation} 
where 
\begin{equation} 
A_{\rm sd} = \frac{\sqrt{-\nabla^2+m^2}} {L_0 K_1 \!\left( L_0\sqrt{-\nabla^2+m^2} \right)}. 
\label{ActionOperator} 
\end{equation} 
Equations (\ref{treinta}) and (\ref{ActionOperator}) provide the sought--for realisation of Padmanabhan's duality principle, now in field--theoretic terms.

The action (\ref{treinta}) differs fundamentally from conventional higher--derivative field theories. In local effective field theories one typically encounters finite or infinite polynomial expansions in powers of the d'Alembertian operator. By contrast, the selfdual action contains the nonpolynomial operator 
$K_1 \!\left( L_0\sqrt{-\nabla^2+m^2} \right)$, together with the square root of the Klein--Gordon operator. The resulting dynamics is therefore genuinely nonlocal. The nonlocality is not introduced {\it ad hoc}\/; rather, it emerges naturally from the requirement of duality invariance.

\section{Discussion}

A remarkable feature of Padmanabhan's duality principle is that it induces a nonvanishing zero--point length while maintaining, in terms of the spacetime manifold, a continuous and differentiable geometric description. The consequences of a nonvanishing quantum of length $L_0$ have been studied previously in a number of contexts. Thus, {\it e.g.}\/, refs. \cite{PADDY3, PADDY4} have analysed corrections to standard results in quantum field theory due to $L_0>0$. In general relativity, a number of relevant quantum--gravity corrections arising from $L_0>0$ have been reported in refs. \cite{KOTHAWALA1, KOTHAWALA2, PESCI2, PESCI3, PESCI4}; for a review see, {\it e.g.}\/,  ref. \cite{NICOLINI}.

The proposal set forth in this paper must be regarded as an alternative to that made by the authors in previous publications \cite{ISIDRO1, ISIDRO2}, in the following sense. Rather than computing quantum--gravity corrections to existing propagators and action integrals, in this article we set out with the goal of developing, {\it ab initio}\/, the quantum field theory analogue of Padmanabhan's selfdual, point--particle theory \cite{PADDY1, PADDY2}. We looked for an action integral and we came up with an explicit suggestion, namely the action integral (\ref{treinta}). The latter provides a compact realisation of the duality principle in terms of a scalar field on Euclidean $\mathbb{R}^D$. Its main properties are:
\begin{enumerate} 
\item Standard Klein--Gordon theory is recovered smoothly in the limit \(L_0\to0\).
\item The theory is selfdual and nonlocal. 
\end{enumerate} 
Moreover, the corresponding propagator (\ref{Gmomentum}) exhibits some key properties: 
\begin{enumerate} 
\item Ultraviolet modes are exponentially suppressed. 
\item The standard Klein--Gordon pole at $p^2+m^2=0$ is preserved. 
\end{enumerate} 
The continuation of the theory to a Lorentzian metric remains a promising field to explore.

\vskip0.5cm
\noindent
{\bf Acknowledgements}\\
The authors acknowledge support by project MCIN/AEI/10.13039/501100011033 and by ERDF, “A way of making Europe”, under project PID2024-162480OB-I00.

 \end{document}